\documentclass[aps,prl,twocolumn,amsmath,amssymb,superscriptaddress,floatfix]{revtex4-2}

\usepackage{graphicx}	
\usepackage{amsfonts}
\usepackage{bm}         	
\usepackage{tabularx}
\usepackage{color}
\usepackage{dcolumn}		
\usepackage{calc}
\usepackage{xcolor,colortbl}
\usepackage{rotating}
\usepackage[colorlinks,citecolor=blue,linkcolor=blue]{hyperref}
\usepackage{hyperref}

\newcommand{\CNN}{Centre de Nanosciences et de Nanotechnologies, CNRS, Universit\'e Paris-Saclay, 91120 Palaiseau, France}
\newcommand{\SPEC}{SPEC, CEA, CNRS, Universit\'e Paris-Saclay, 91191 Gif-sur-Yvette, France}

\begin{document}

\title{Experimental observation of vortex gyration excited by surface acoustic waves} 

\author{R. Lopes Seeger}
\email{rafael.lopesseeger@cea.fr}
\affiliation{\CNN}
\affiliation{\SPEC}
\author{F. Millo}
\affiliation{\CNN}
\author{G. Soares}
\affiliation{\SPEC}
\author{J.-V. Kim}
\affiliation{\CNN}
\author{A. Solignac}
\affiliation{\SPEC}
\author{G. de Loubens}
\affiliation{\SPEC}
\author{T. Devolder}
\affiliation{\CNN}

\date{\today}

\begin{abstract}
We report experiments using magnetic resonance force microscopy to investigate the dynamics of magnetic vortices in sub-micrometer CoFeB disks grown on a piezoelectric substrate. We compare the vortex gyration excited inductively by microwave magnetic fields and through magnetoelastic coupling via surface acoustic waves (SAWs). Based on the device geometry and modeling performed, we find that the dominant mechanism for SAW-driven excitation is through lattice rotations, which generate an effective field localized at the vortex core. This contrasts with the magnetoelastic strains typically assumed in similar experiments. Moreover, we demonstrate that this magnetorotation torque can be tuned by applying a perpendicular magnetic field. Unlike spin wave excitations, nontrivial spin textures, such as vortices, effectively couple to magnetorotation torques, making them valuable probes for studying such phenomena.
\end{abstract}

\maketitle

Driving magnetization dynamics with surface acoustic waves (SAWs) has garnered recent attention due to promising applications in magnonics and microwave acoustics~\cite{Weiler2011, Thevenard2014, Sasaki2017, Puebla2020}. Particular focus has been directed toward coupling spin waves with SAWs, since the nonlinear and nonreciprocal properties of the former can potentially be conferred to the latter~\cite{Verba2019, Hernandez2020, Xu2020, Millo2023, Kuess2024}, opening up new device possibilities such as acoustic isolators. Besides ferromagnets, recent advancements have extended SAW-driven magnetization dynamics to multiferroics~\cite{Sasaki2019}, synthetic antiferromagnets \cite{Matsumoto2022}, and layered antiferromagnets~\cite{Lyons2023}.

Most studies to date have only focused on the role of the magnetoelastic strain induced by SAWs, $\epsilon_{ij} = (\partial u_i/\partial x_j + \partial u_j/\partial x_i)/2$, where $u$ denotes the lattice displacement and $i,j$ the different Cartesian coordinates. For example, longitudinal strain of the form $\epsilon_{xx}$ produces a magnetoelastic ``tickle'' field that can induce the motion of magnetic vortex cores in thin-film disks, but only with a static magnetic field applied in the film plane which displaces the equilibrium position of the core~\cite{Koujok2023}. This approach was recently implemented using a time-varying-strain in the MHz range to drive the vortex gyrotropic mode~\cite{iurchuk2023excitation}. Alternatively, additional nonzero lattice deformations can also appear in real piezoelectric materials with different symmetries to longitudinal strain~\cite{Yamamoto2023, Seeger2024}. In particular, the rotational motion of the lattice within the magnetic material causes reorientation of the surface normal direction, which couples to the magnetization via magnetic anisotropy fields~\cite{Maekawa1976}. This so-called \emph{magnetorotation} contribution (resulting in a ``rolling'' field), $\omega_{ij} = (\partial u_i/\partial x_j - \partial u_j/\partial x_i)/2$, is active in materials with vanishing magnetostriction. Here, we demonstrate experimentally that magnetic vortices are sensitive probes for magnetorotation torques when the vortex radial symmetry is preserved. In this regime, steady-state gyration can be induced by SAWs without an in-plane applied fields to displace the vortex core from the disk center.

\begin{figure} 
\includegraphics[width=5.7cm]{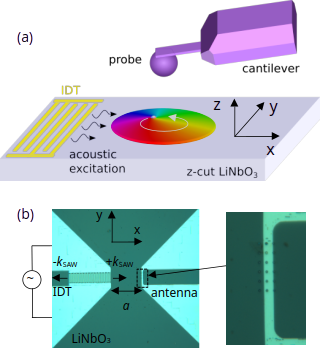}
\caption{
(a) Sketch of the MRFM experiment to detect the vortex gyration induced by SAWs. An IDT excites SAWs on a piezoelectric substrate towards a disk in the vortex state (the colors encode the in-plane magnetization component). The disk is placed $a = 104 ~\mu$m away from the IDT. (b) The setup includes a top antenna for inductive excitation, shown in an optical image. The inset is a zoomed view of the antenna's constriction with several CoFeB microdisks. } 
\label{sketch}
\end{figure}

Figure~\ref{sketch}(a) presents a schematic illustration of the experimental setup.  
We study the influence of SAW propagation through a CoFeB disk positioned on a $Z$-cut LiNbO$_3$ piezoelectric substrate. The disk has a thickness of $d_\mathrm{FM} = 34$ nm and a diameter of $2L = 800$ nm. Resonant SAWs with a frequency of 0.86 GHz and a wavelength of 4.5 $\mu$m, corresponding to the second harmonic order, are generated by applying an rf voltage on the interdigital transducer (IDT) with a periodicity 9 $\mu$m. They propagate on the surface of the piezoelectric substrate~\cite{SM} along the crystalline $X$-axis of LiNbO$_3$, which coincides with the $x$-axis shown in Fig.~\ref{sketch}(a). When the SAWs impinge on the CoFeB disk, they induce lattice vibrations in the ferromagnet which give rise to additional contributions to the effective magnetic field through the magnetoelastic and magnetorotation coupling. To provide a point of comparison, a microwave antenna [with a width of $w_\textrm{ant} = 4~\mu$m, Fig. \ref{sketch}(b)] is also placed on top of the disk to excite magnetization dynamics inductively with rf magnetic fields. For both excitation methods, detection of the magnetization dynamics is performed with a magnetic resonance force microscope (MRFM)~\cite{Klein2008, Loubens2009, SM}, providing unequivocal proof of the magnetoacoustic origin of the detected excitation.

\begin{figure} 
\includegraphics[width=8.5cm]{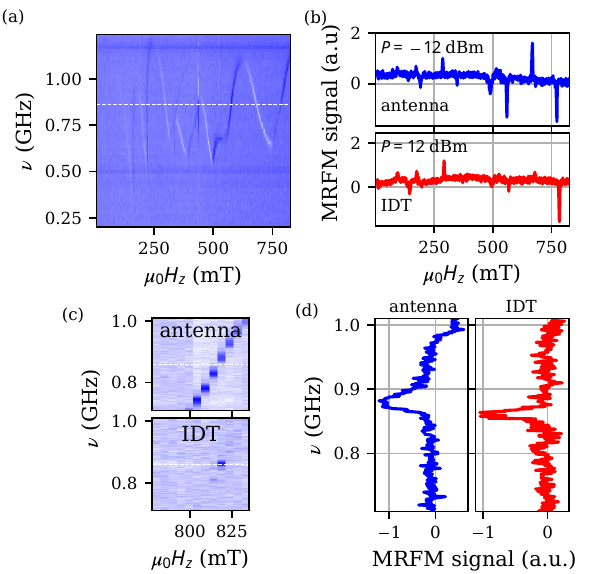}
 \caption{
Comparison of vortex gyrotropic excitation using microwave inductive and magnetoacoustic fields. (a) Vortex gyration frequency versus perpendicular magnetic field using the broadband inductive antenna, with MRFM signal amplitude as contrast. (b) Gyrotropic spectroscopy as a function of perpendicular field with excitation from either the antenna or IDT at 0.86 GHz (2nd harmonics of SAW). (c) Frequency scans comparing inductive and acoustic vortex excitations. (d) Line cuts of the data in (c) under $\mu_0 H_z = $ 820 mT.
}
\label{Experiment}
\end{figure}

MRFM spectroscopy of the magnetic vortex dynamics induced by an in-plane microwave field is presented in Fig.~\ref{Experiment}(a). By sweeping the amplitude of the perpendicular magnetic field, $H_z$, at fixed microwave frequency, we observe an erratic variation of the gyration frequency of the vortex, ranging from 0.4 to 1.2 GHz, instead of the expected linear variation \cite{Loubens2009}, from $\nu_G = 5/(9\pi) \gamma \mu_0 M_s (d_{\mathrm{FM}}/L) \simeq 0.72~\textrm{GHz}$ at zero field to 
1.44 GHz close to saturation 
\footnote{$H_\textrm{sat} \simeq M_s$ is the saturation field of the disk, $\mu_0 M_s \simeq 1.7$ T the CoFeB saturation magnetization, and $\gamma/(2 \pi) \simeq 28$ GHz/T its gyrotropic ratio.}. 
Similar results were obtained on another CoFeB disk~\cite{SM}. 
This behavior is attributed to the strong pinning of the vortex core by grain defects in the CoFeB disk, a phenomenon that has also been reported for other material systems~\cite{Compton2006, shreya_granular_2023}. In Fig.~\ref{Experiment}(b), we show the MRFM spectrum of the SAW-driven vortex gyration. 
We observe similar features in gyrotropic response in the two cases, although there are small differences in the relative amplitudes of spectral features that might arise from the distinct physical mechanisms behind the effective fields generated. 


We also conducted frequency scans at fixed perpendicular field with both the microwave antenna and the IDT [Figs.~\ref{Experiment}(c) and \ref{Experiment}(d)]. While a broadband excitation is observed for field-driven dynamics with the microwave antenna, SAW-driven excitations are only effective at the resonance of the IDT with the SAWs, namely around $\nu \approx 0.86$ GHz. As such, a large response in the IDT-driven case is only observed when the gyration frequency is appropriately tuned by $H_z$, which for the system studied is around 820 mT. Note that the spectral response under IDT driving is narrower than the response under field excitation, i.e., 15.7 vs. 30.9 MHz, respectively. This is because the measured response under IDT-driving is primarily determined by the SAW resonance, which is narrower than the overall gyrotropic response of the vortex. This observation lends further support to magnetoacoustic torques being responsible for vortex excitation in the IDT-driven case.



\begin{figure}
\includegraphics[width=8.45cm]{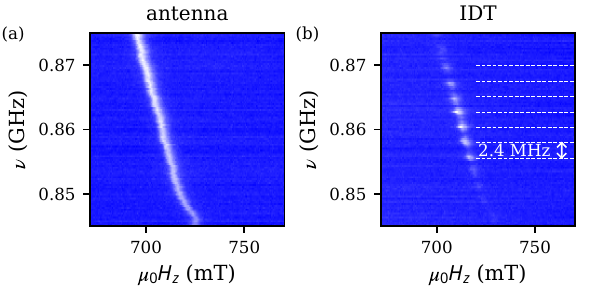}
\caption{High-frequency resolution gyrotropic spectroscopy. Perpendicular magnetic field dependence for an excitation employing the antenna (a) or the IDT (b). In the latter case, oscillations with a period of $\approx 2.4$ MHz appear due to standing acoustic modes in the LiNbO$_3$ substrate. 
}
\label{FineStruct}
\end{figure}

To gain further insight into the magnetoacoustic signal, we performed experiments with higher frequency resolution around the SAW resonance, as shown in Fig.~\ref{FineStruct}.  
For the antenna excitation, the response remains nearly constant over the frequency interval considered, while strong oscillations in the MRFM signal are observed for the IDT excitation. These oscillations arise from standing acoustic modes formed along the SAW propagation direction, which are caused by the reflection from the edge of the substrate~\cite{An2020}. Traveling SAWs along the LiNbO$_3$ surface can indeed propagate over millimeter distances~\cite{Casals2020}. The sample length along the SAW propagation direction is approximately $\ell=1.8$ mm, which for the typical SAW velocity in LiNbO$_3$ ($v = 3900$ m/s) results in a frequency splitting $v/\ell$ of around 2.2 MHz between consecutive standing modes, in good agreement with the oscillation period observed in Fig.~\ref{FineStruct}(b)~\cite{SM}. The consequence of standing acoustic modes is that the elastic power injected by the IDT is effectively spread out over the length $\ell$ of the substrate, in contrast to the more localized power of the microwave field generated by the inductive antenna. The observation of such a fine structure in the SAW signal unambiguously points towards the acoustic origin of the vortex gyration induced by the IDT excitation.

\begin{figure} 
\includegraphics[width=7.1cm]{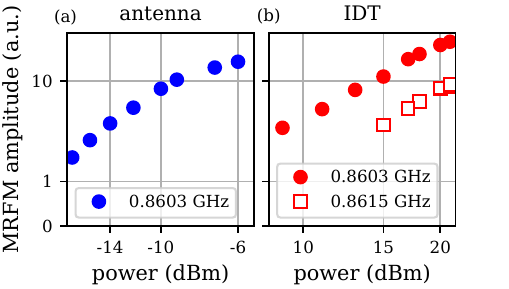}
\caption{MRFM signal amplitude as a function of the excitation power. Logarithmic-scale dependence as measured with (a) inductive and (b) acoustic excitation as measured for different microwave frequencies. (Inset) Selected MRFM spectra. The experiments are performed in the same range of perpendicular field as in Fig. \ref{FineStruct}. 
} 
\label{Power}
\end{figure}

Let us now turn to the magnitude of the effective excitation field generated by the magnetoacoustic coupling. Fig.~\ref{Power} compares the dependence of the MRFM signal amplitude at the gyration frequency, for both inductive and acoustic excitations, on the input power, $P$. 
We observe a monotonic increase in the gyration signal as a function of $P$. At low excitation powers a symmetric resonance peak is observed, which is consistent with a regime of linear dynamics. At higher powers, we observe hallmarks of nonlinearity in the spectral response, such as a distortion in the peaks [inset in Fig. \ref{Power}(b)], as reported in previous works~\cite{Pigeau2010, Drews2012}. The similar power dependence between the two excitation methods used suggests that the propagating SAWs also exert torques that are capable of exciting large precession amplitudes, albeit requiring 24 dB more electrical power. However, this figure belies a higher efficiency of the magnetoacoustic coupling, since a more accurate comparison should account for the effective magnetic volumes driven by the two sources; we estimate this to be $20~ \textrm{log}_{10}(\ell / w_\textrm{ant}) \approx 53~\textrm{dB}$. The energy efficiency of magnetoacoustic excitation is, in fact, three orders of magnitude greater than that of inductive excitation. 
%
%

\begin{figure}
\includegraphics[width=8.55cm]{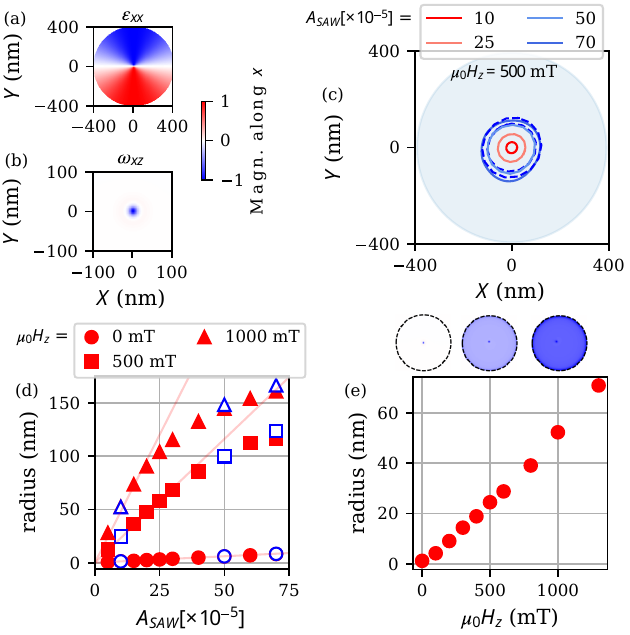}
\caption{
Normalized magnetoacoustic field ($b_{\textrm{eff}}$) distribution over the vortex due to (a) lattice strain $\epsilon_{xx}$ and (b) rotation $\omega_{xz}$ at zero applied field, with color coding indicating the $x$-direction field. (c) Vortex core trajectories for selected deformation amplitudes $A_{\textrm{SAW}}$ 
at the gyrotropic frequency ($\mu_0 H_z$ = 500 mT). Bold lines represent trajectories driven by $\epsilon_{xx}$, $\epsilon_{zz}$, and $\omega_{xz}$; dashed lines show those driven by $\omega_{xz}$ alone. (d) Trajectory radius as a function of $A_{\textrm{SAW}}$ for different applied fields, with open symbols showing $\omega_{xz}$-only effects. (e) Trajectory radius versus applied field $\mu_0 H_z$ for $A_{\textrm{SAW}} = 10 \times 10^{-5}$. (Inset) Normalized rolling fields distribution over the vortex with different applied fields, with dashed circles indicating the disk edge.
}
\label{Simul}
\end{figure}

To gain more insight into the role of the magnetoacoustic torques, we performed micromagnetics simulations of the vortex dynamics using the \texttt{MuMax3} code~\cite{Vansteenkiste2014}. We model the SAW impinging on the CoFeB disk as a plane wave with frequency $\nu$ and deformation amplitude $A_{\textrm{SAW}}$ for the strain components $\epsilon_{xx}$ and $\epsilon_{zz}$, as well as for the rotation component $\omega_{xz}$~\cite{SM}. Note that the shear strain $\epsilon_{xz}$ exactly vanishes at the surface between the piezoelectric substrate and the magnetic disk. These torques correspond to the non-zero lattice strain and rotation components of a Rayleigh SAW traveling in an isotropic substrate mimicking the Z-cut of LiNbO$_3$~\cite{Seeger2024}. We first examine the effect of the longitudinal strain alone. As discussed in Refs. ~\onlinecite{Ostler2015} and ~ \onlinecite{Koujok2023}, $\epsilon_{xx}$ alone cannot displace the vortex core as its corresponding effective field is antisymmetric about the disk center [Fig. \ref{Simul}(a)], resulting in a zero net torque acting on the vortex core. This issue can be circumvented by applying an additional static in-plane magnetic field, which shifts the equilibrium position of the core away from the center. This field breaks the axial symmetry of the static vortex, resulting in a finite torque from the strain component that can subsequently drive the core dynamics~\cite{Koujok2023, iurchuk2023excitation}. In experiments, if the applied field is not perfectly aligned in the out-of-plane direction, it will result in an in-plane field projection that may shift the equilibrium position away from the disk center.
We next consider the magnetorotation term, $\omega_{xz}$, which results in the effective field profile shown in Fig.~\ref{Simul}(b). In contrast to the tickle field, the effective rolling field associated with $\omega_{xz}$ is non-vanishing and highly localized to the core. This term therefore exerts a net torque on the core irrespective of its position in the disk, thereby providing a means of initiating gyration from equilibrium and at zero in-plane applied field.

Figure~\ref{Simul}(c) shows the simulated core trajectories for selected values of $A_{\textrm{SAW}}$, with $\mu_0 H_z =$ 500 mT. The trajectories are circular for moderate deformation amplitudes but become distorted and elliptical for large values of $A_{\textrm{SAW}}$. To highlight the role of the tickle and rolling fields, the dashed line in Fig.~\ref{Simul}(c) shows the circular trajectory obtained with only the $\omega_{xz}$ component at $A_{\textrm{SAW}} = 50 \times 10^{-5}$ and $70 \times 10^{-5}$. Fig.~\ref{Simul}(d) shows the variation of the average radius of gyration as a function of $A_{\textrm{SAW}}$ for three different values of $\mu_0 H_z$. The orbital radius increases linearly with $A_{\textrm{SAW}}$ up to a certain value. Within this range of magnetoacoustic torques, the gyration frequency remains largely independent of $A_{\textrm{SAW}}$, with nonlinear frequency shifts and noncircular orbits appearing once stronger torques are applied. For $\mu_0 H_z = $ 500 mT, for example, the linear behavior is observed for $A_{\textrm{SAW}} < 25 \times 10^{-5}$. The closed symbols in Fig.~\ref{Simul}(d) correspond to cases in which both the tickle and rolling fields are applied simultaneously, which show that the strain terms only provide a negligible correction in the orbital radius (and only beyond linear regime), meaning that the rotational contribution alone is sufficient to induce steady state gyration of the vortex core. For that reason, whether the SAWs couple to the gyrotropic mode is determined by the effective magnetoacoustic fields acting on the core at its equilibrium position. The rotation term acts to displace the vortex core from equilibrium at the disk center, but as it gyrates the couple of the strain terms averages to zero.

Figure~\ref{Simul}(e) presents the dependence of the orbital radius on the applied perpendicular field with $A_{\textrm{SAW}} = 10 \times 10^{-5}$. With this SAW amplitude, the core trajectory is circular over the full range of studied fields. 
The perpendicular field acts to tilt the magnetic moments of the curling vortex texture out of the plane toward the field, resulting in a vortex cone state with a larger core~\cite{Ivanov2002, Loubens2009}. 
The larger orbital radius with increasing $\mu_0 H_z$ arises from a larger rolling field that appears across the full canted structure, rather than being only localized at the core at $\mu_0 H_z =$ 0 [inset in Fig. \ref{Simul}(e)].

Further insight into the role of the different magnetoacoustic torques can be gained through the rigid-core approximation. In this approach, the magnetization dynamics is determined solely by the motion of the core, whose coordinates $\mathbf{X}=[X(t), Y(t)]$ evolve according to the Thiele equation~\cite{Guslienko2008, SM},
\begin{equation}
\mathbf{G} \times \frac{d \mathbf{X}}{dt} + \alpha \mathbf{D} \cdot \frac{d \mathbf{X}}{dt} = -\frac{\partial U}{\partial \mathbf{X}},
\label{eq:Thiele}
\end{equation}
where $\mathbf{G} $ is the gyrovector, $\mathbf{D} $ is the damping tensor, and $U$ represents the effective potential energy seen by the vortex core. Without external driving, $U = U_0(\| \mathbf{X} \|^2)$ is a central potential governed by the confining effects of the disk, resulting in damped gyrotropic motion of the vortex core around the disk center. To illustrate the mechanics of this approach, consider first the more familiar example of a spatially uniform in-plane microwave field applied along $x$, which enters the Zeeman term (assuming $\mu_0 H_z = 0$)
\begin{equation}
U_\mathrm{Z}(t) = -\mu_0 M_s \int dV \; h_{x}(t) \, m_x(\mathbf{r},t).
\label{eq:rffield}
\end{equation}
Because the dominant contribution to the integral comes from the region outside the core, i.e., where $m_z(\mathbf{r}) \simeq 0$, we can take $m_x(\mathbf{r}) \simeq \cos\Phi_0$, where $\Phi_0 = \tan^{-1}\left[ y-Y/(x-X) \right] \pm \pi/2$, which gives $U_\mathrm{Z} = \mp \mu_0 M_s \pi d_\mathrm{FM} L h_{x} Y $ and results in the time-dependent force 
\begin{equation}
\mathbf{F}_\mathrm{Z}(t) = -\frac{\partial U_{\mathrm{Z}}}{\partial \mathbf{X}} = \left(0, \pm \mu_0 M_s \pi d_\mathrm{FM} L h_{x}(t) \right).
\end{equation}
This represents a ``volumic'' contribution and is independent of the core size and position within the approximation used.

We next discuss the time-dependent magnetoelastic term
\begin{equation}
U_\mathrm{me}(t) = B_1 \int dV \; \epsilon_{ii}(t) \, m_i(\mathbf{r},t)^2,
\label{eq:meeng}
\end{equation}
with $i = x,z$, which acts like a uniaxial anisotropy in the film plane ($\epsilon_{xx}$) or perpendicular to the plane ($\epsilon_{zz}$). The $\epsilon_{xx}$ term is also volumic and depends on the in-plane magnetization outside the core. By using the same approximation as for the Zeeman term, we find to lowest order for small core displacements  $U_{\mathrm{me},xx} = (\pi B_1 d_\mathrm{FM})(2-X^2+Y^2)/4$, which represents a hyperbolic paraboloid potential. In contrast, the $\epsilon_{zz}$ term only depends on the perpendicular component of the magnetization, $m_z(\mathbf{r},t)$, which is only nonzero in the core region. While this term can excite radial spin wave modes of the vortex state~\cite{Ivanov2005, Vogt2011, Taurel2016}, it gives no contribution to Eq.~(\ref{eq:Thiele}), since the integral of $m_z(\mathbf{r},t)^2$ over the disk is constant and therefore independent of the core position, $U_{\mathrm{me},zz} = \mathrm{const}$. The resulting magnetoelastic force is then
\begin{equation}
\mathbf{F}_\mathrm{me} = -\frac{\partial U_{\mathrm{me},xx}}{\partial \mathbf{X}} = \frac{\pi B_1 d_\mathrm{FM}}{2} \epsilon_{xx}(t) \left(X, -Y \right),
\end{equation}
which vanishes when the core is at the disk center, $\mathbf{X} = \mathbf{0}$, as discussed earlier.

Consider now the time-dependent magnetorotation 
\begin{equation}
U_\mathrm{mr}(t) = \mu_0 M_\mathrm{s}^2 \int dV \; \omega_{xz}(t) \, m_x(\mathbf{r},t) m_z(\mathbf{r},t),
\label{eq:mreng}
\end{equation}
%
Unlike $\epsilon_{xx}$, which is driven by the in-plane magnetization, or $\epsilon_{zz}$, which is dominated by the core center, the magnetorotation term is driven by the boundary or ``waist'' of the core at which $\sin\Theta_0 \cos\Theta_0$ is largest, where $\Theta_0$ represents the polar angle of the magnetization. Compared to the Zeeman term, the core ``waist'' appears thus like an additional convolution kernel of $\cos\Phi_0$ in $U_\mathrm{mr}$. For small displacements of the core, the force can be expressed in terms of the static core profile $\Theta_0=\Theta_0(r)$ as 
\begin{equation}
\mathbf{F}_\mathrm{mr} = -\frac{U_\mathrm{mr}}{\partial \mathbf{X}} = \left(0, \mp \pi \mu_0 M_\mathrm{s}^2 d_\mathrm{FM} \omega_{xz} f(\Theta_0) \right),
\end{equation}
where $f(\Theta_0)$ is a structure factor
\begin{equation}
f(\Theta_0) = \int dr \left( r \frac{\partial \Theta_0}{\partial r} \cos2\Theta_0 + \frac{\sin{2\Theta_0}}{2} \right).
\end{equation}
The integral can be evaluated with the Usov \emph{ansatz} to give $f(\Theta_0) = (1-\log{2})b$, where $b$ is the core radius. For $\mu_0 H_z >$ 0, $m_z$ is nonzero outside the core which further enhances $f(\Theta_0)$.

Our results demonstrate that nonuniform spin textures like magnetic vortices can be useful for probing magnetorotation coupling like $\omega_{xz}$. Beyond offering new avenues towards SAW-based microwave magnetoacoustics, our findings also suggest that other nontrivial textures, such as domain walls and skyrmions, could also exhibit behaviors that are not accessible through magnetoelastic strains alone.

\begin{acknowledgments}
%
This work was supported by the Agence Nationale de la Recherche (ANR) under Contract No. ANR-20-CE24-0025 (MAXSAW), the ``Investissements d'Avenir'' program LabEx NanoSaclay, under Contract No. ANR-10-LABX-0035 (SPICY), and the France2030 program PEPR SPIN, under Contract Nos. ANR-22-EXSP-0008 and ANR-22-EXSP-0004.
\end{acknowledgments}

\bibliography{scibib}

\clearpage

\end{document}